\documentclass[letterpaper, 10 pt, conference]{ieeeconf}
\IEEEoverridecommandlockouts
\usepackage{cite}
\usepackage{amsmath,amssymb,amsfonts}
\usepackage{algorithmic}
\usepackage{graphicx}
\usepackage{textcomp}
\usepackage{cite}
\usepackage{amsmath,amssymb,amsfonts}
\usepackage{algorithmic}
\usepackage{graphicx}
\usepackage{textcomp}
\usepackage{multicol}
\usepackage{multirow}
\usepackage{bm}
\usepackage{color}
\newcommand*\xoverline[2][0.75]{%
    \sbox{\myboxA}{$\m@th#2$}%
    \setbox\myboxB\null% Phantom box
    \ht\myboxB=\ht\myboxA%
    \dp\myboxB=\dp\myboxA%
    \wd\myboxB=#1\wd\myboxA% Scale phantom
    \sbox\myboxB{$\m@th\overline{\copy\myboxB}$}%  Overlined phantom
    \setlength\mylenA{\the\wd\myboxA}%   calc width diff
    \addtolength\mylenA{-\the\wd\myboxB}%
    \ifdim\wd\myboxB<\wd\myboxA%
       \rlap{\hskip 0.5\mylenA\usebox\myboxB}{\usebox\myboxA}%
    \else
        \hskip -0.5\mylenA\rlap{\usebox\myboxA}{\hskip 0.5\mylenA\usebox\myboxB}%
    \fi}
\makeatother

\usepackage{comment}
\usepackage{siunitx}
\usepackage{url}

\title{NeAS: 3D Reconstruction from X-ray Images using Neural Attenuation Surface}

\author{Chengrui Zhu$^{1}$, Ryoichi Ishikawa$^{1}$,,Masataka Kagesawa$^{1}$, Tomohisa Yuzawa$^{1,2}$,  Toru Watsuji$^{2}$, and Takeshi Oishi$^{1}$,
\thanks{$^{1}$The authors are with The Institute of Industrial Science, The University of Tokyo, Japan. Emails:
		{\tt\small \{zhucr, Ishikawa, kagesawa, yuzawa-tom,  oishi\}@cvl.iis.u-tokyo.ac.jp}}%
\thanks{$^{2}$The authors are Air Water Inc., Minami Semba, Chuo-ku, Osaka, Japan. Emails:{\tt\small watsuji-tor@awi.co.jp}}%
}

\begin{document}
\maketitle

\begin{abstract}
Reconstructing three-dimensional (3D) structures from two-dimensional (2D) X-ray images is a valuable and efficient technique in medical applications that requires less radiation exposure than computed tomography scans.
Recent approaches that use implicit neural representations have enabled the synthesis of novel views from sparse X-ray images.
However, although image synthesis has improved the accuracy, the accuracy of surface shape estimation remains insufficient.
Therefore, we propose a novel approach for reconstructing 3D scenes using a Neural Attenuation Surface (NeAS) that simultaneously captures the surface geometry and attenuation coefficient fields.
NeAS incorporates a signed distance function (SDF), which defines the attenuation field and aids in extracting the 3D surface within the scene.
We conducted experiments using simulated and authentic X-ray images, and the results demonstrated that NeAS could accurately extract 3D surfaces within a scene using only 2D X-ray images.
\end{abstract}

% keywords can be removed
%\keywords{3D Reconstruction \and Implicit Neural Representation \and X-ray Image \and Neural Rendering \and Surface Extraction.}

\section{Introduction}
\label{sec:introduction}
%\IEEEPARstart{T}{hree}
Three-dimensional (3D) imaging of anatomical structures is helpful in clinical applications, such as diagnosis, surgery planning, and postoperative follow-up.
The typical approach to obtain such 3D information is computed tomography (CT). 
%and magnetic resonance imaging (MRI).
CT provides 3D, high-resolution, and high-contrast images that help detect bone and blood vessel abnormalities, infections, tumors, and other symptoms.
However, CT requires the capture of numerous slices or images to achieve high-resolution modeling, which exposes patients to high-level ionizing radiation. 
%and incurs significant costs and time in MRI. 

Among the CT methods, cone beam CT (CBCT) offers a relatively low radiation dose, and enhancements have been made to the accuracy and resolution of 3D reconstruction~\cite{yeung2019novel, alaei2015imaging, liu2023review}. 
There exists a trade-off among the number of images acquired and the performance of CBCT. 
Since acquiring fewer images inevitably introduces noise and artifacts into the reconstructed CT volume, mitigating these issues can help reduce radiation exposure. 
Consequently, extensive research in recent years has focused on integrating deep learning methods for CT reconstruction~\cite{biguri2024advancing, amirian2024artifact}.

Recent studies have employed implicit neural representations (INRs) to reconstruct CT volumes from sparse X-ray images~\cite{naf, neat, snaf, sax_nerf}.
Over the past several years, INRs such as neural radiance fields (NeRF) have gained popularity as a method for representing 3D scenes~\cite{nerf, zhou2023inf, kerbl3Dgaussians}.
By encoding scenes into the weights of neural networks via differentiable rendering techniques, INRs enable the high-precision reconstruction of 3D scenes with complex lighting and material properties.
In the context of X-ray imaging, the neural attenuation field (NAF) allows for the estimation of the attenuation coefficient at each point~\cite{naf}.
Owing to its precise reconstruction capabilities, CT volumes can be generated from a relatively small number of images. 

However, these INR-based methods do not accurately reconstruct the surface geometry of the objects in the scene.
When extracting a surface from a CT volume, mesh data is produced from voxel values using techniques such as the Marching Cubes algorithm.
If the volume lacks sufficient resolution, aliasing issues can arise.
Moreover, since INR-based methods learn attenuation fields by maintaining consistency between the input and rendered images, the problem of 3D-2D ambiguity is unavoidable.
This problem is particularly pronounced in X-ray imaging, where the visual perspective shifts significantly with changes in viewpoint. 
%because of the ambiguity in the 3D--2D projections. 

Therefore, we introduced Neural Attenuation Surface (NeAS) to acquire accurate 3D shapes from X-ray images within the INR framework.
NeAS utilizes the signed distance function (SDF)~\cite{neus} to describe the target surface that bounds the neural attenuation fields.
Because a single set of attenuation fields and an SDF cannot represent the boundary surfaces between several different materials, such as bones and muscles, we also proposed an architecture for cases involving multiple materials.
In addition, since precisely calibrated systems are extremely costly, the proposed method employs a pose refinement approach to account for errors in extrinsic parameters. 
%: two materials in this paper. 
%Moreover, to address the issue of low training and inference speeds, we incorporate the multiresolution hash-encoding method~\cite{instantngp} into the proposed method.

The contributions of this paper can be summarized as follows: 
\begin{itemize}
\item We propose an INR-based framework that integrates SDF and NAF by introducing the surface boundary function (SBF), enabling accurate 3D surface reconstruction from X-ray images.
\item We also introduce an approach that simultaneously estimates multiple object surfaces, such as bone and skin, using multiple attenuation fields. 
\item We demonstrate that incorporating a robust pose refinement method with frequency regularization allows for more accurate estimation of attenuation and signed distance fields. 
\end{itemize}
We further show that our approach enhances performance in a novel view synthesis task and successfully extracts surfaces even in scenes containing multiple materials.
%In addition, we provide an alternative approach for bone density estimation by analyzing the attenuation field obtained using our framework.

%%%%%%%%%%%%%%%%%%%%%%%%%%%%%%%%%%%%%%%%%%%%%
\section{Related work}
\label{sec:relatedwork}
In this section, we begin with a brief review of previous studies on 3D surface reconstruction from X-ray images. 
Since CBCT has a long history and numerous survey papers have been published~\cite{yeung2019novel, alaei2015imaging, liu2023review, biguri2024advancing, amirian2024artifact}, we will focus on introducing surface estimation methods within the field of computer vision. 
Following this, we provide an explanation of INR methods relevant to this study and discuss INR-based 3D reconstruction from X-ray images.

\subsection{3D Surface Reconstruction from X-ray Images}

Most traditional 3D reconstruction approaches rely on the deformation models~\cite{review, 4106750, 4804738}.
These methods typically require 3D template models of the target, which are subsequently transformed based on input X-ray images.
The statistical shape model (SSM)~\cite{SSM}~\cite{lamecker2006atlas} and statistical shape and intensity model (SSIM)~\cite{SSIM}~\cite{ssim1} are frequently employed to characterize such deformation models.
These models establish statistical representations of the shape geometry and variances, resulting in a reduced number of parameters for optimization compared with general-purpose models.

The effectiveness of deep learning has been demonstrated in various tasks, including 3D reconstruction from X-ray images.
CNN excels in object detection, and SSM is deformed using detected feature points and boundaries for 3D reconstruction~\cite{cnn1, cnn+ssm}.
Hybrid deep neural network~\cite{liver} that uses CNN for feature extraction and graph convolutional network (GCN) for model transformation has also shown good performance.
Meanwhile, an end-to-end 3D reconstruction approach~\cite{e2e} eliminates the need for prior 3D knowledge, where the CNN directly outputs a 3D reconstruction model.
GAN~\cite{med-gan} can also be applied for 3D generation based on given 2D X-ray images.
However, these methods often struggle with the absence of a prior model or a lack of training data.

\subsection{Implicit Neural Representation }

INR is a method for representing 3D scenes with neural networks, as opposed to explicit representations, such as point clouds and triangle meshes.
%, which discretely define the surface geometry.
NeRF~\cite{nerf} demonstrated promising results in novel view synthesis by optimizing a neural network through ray sampling and volume rendering.
However, the original NeRF faces challenges such as slow training and inference speeds.
Additionally, NeRF requires a substantial number of images, such as several hundred images, to achieve sufficient quality in the synthesized images, posing a significant limitation.

Recently, more research has attempted to overcome these limitations of the NeRF.
The remainder of this section describes methods for improving the NeRF that are related to the proposed method.

\subsubsection{Acceleration}
Various methods are aimed at improving NeRF's training speed.
The general approach involves storing 3D feature embeddings in an explicit spatial structure combined with a compact decoder network to query the density and color of a sampled point.
Point-NeRF~\cite{pointnerf} adopts a point cloud to embed points, whereas TensoRF~\cite{tensorf} uses a voxel grid with low-rank decomposition.
Instant-NGP~\cite{instantngp} implements a multiresolution hash map, which significantly reduces the memory requirements of the grid representation.
Plenoxels~\cite{plenoxels} innovatively eliminates the decoder network and acquires color directly from the spherical harmonics.
The proposed method also uses hash encoding to query sampled points.

\subsubsection{Geometry-centric approach}
Although NeRF excels in rendering high-quality 2D images, it struggles to reconstruct the 3D geometry accurately.
Thus, recent efforts have focused on enhancing the geometric consistency of surfaces within a scene.
For instance, UniSurf~\cite{unisurf} simultaneously learns implicit surfaces and NeRFs using an occupancy field to represent the surfaces.
Meanwhile, NeuS~\cite{neus} employs implicit surfaces and represents them using an SDF.
SDF is utilized in the volume-rendering process and enables the extraction of surfaces at their zero-level sets.
Neus2~\cite{neus2} employs hash encoding to accelerate the training and inference processes and to extend the representation capability to dynamic scenes.
Our work also incorporates SDF for surface learning, but uniquely applies it to X-ray images, differing from the approach used for RGB images.

\subsubsection{Pose refinement}

Accurate camera parameters are crucial for successful implementation of NeRF.
Therefore, several studies have focused on optimizing the camera parameters while training the NeRF.
For example, BARF~\cite{barf} introduced a coarse-to-fine strategy for positional encoding of camera poses.
Meanwhile, SCNeRF~\cite{SCNeRF} implemented geometric loss to enhance the optimization of camera poses and lens distortion.
GRAF~\cite{GRAF} learned poses from a random distribution.
NoPe-NeRF~\cite{nope} utilized mono-depth maps to simultaneously constrain relative frame poses and regularize NeRF's geometry.

To mitigate inaccurate camera parameters when taking X-ray photographs, our work also incorporates pose refinement during training and optimizes the camera parameters concurrently with the attenuation field.

\subsection{INR-Based X-Ray Imaging}
Several studies used implicit representations for CT volume reconstruction.
For instance, NAF~\cite{naf} learns a neural attenuation field for CBCT reconstruction using a hash encoder similar to that in~\cite{instantngp} for efficient training. 
Meanwhile, SNAF~\cite{snaf} integrates a deblurring network to address the sparse-view problem inherent in NAF.
NeAT~\cite{neat} uses an octree grid structure to store 3D feature embeddings coupled with a compact multilayer perceptron (MLP) for decoding. 
SAX-NeRF uses Transformer to enable the recovery of detailed structural information during CT reconstruction~\cite{sax_nerf}. 

As mentioned above, these approaches face challenges in accurately reconstructing 3D surface geometries.
This is similar to the scenario seen between NeRF and geometry-centric methods such as NeuS~\cite{neus, neus2} in the visible spectrum. 
Therefore, 
%using appropriate methods, 
we designed a framework to extract surfaces and enhance geometric consistency in the neural attenuation field. 

\begin{figure*}[t]
\begin{center}
\includegraphics[width=0.9\linewidth]{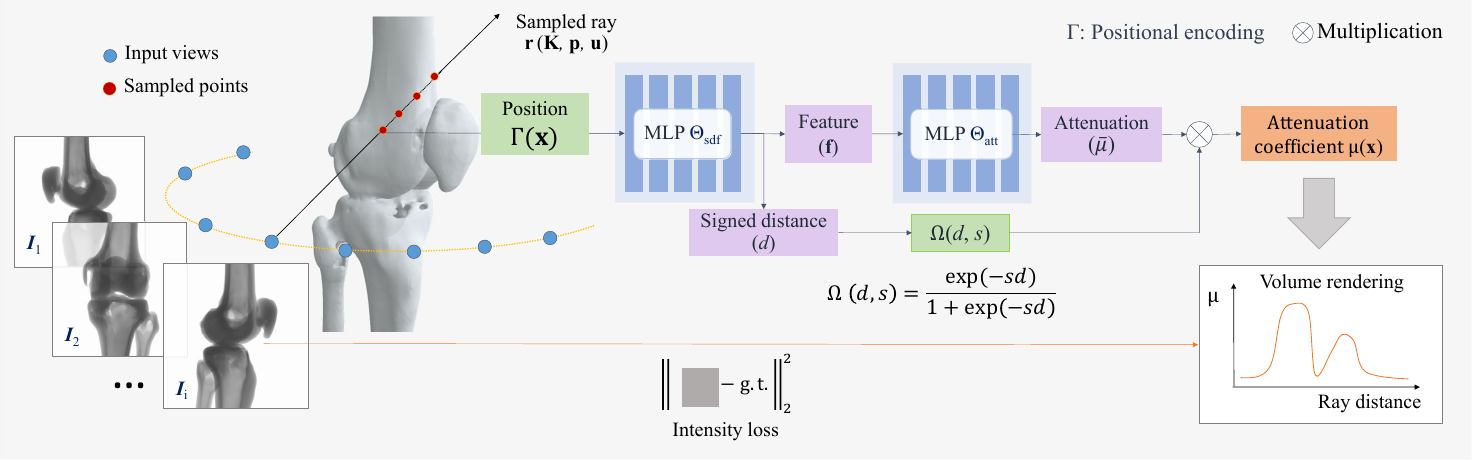}
\caption{Overview of our framework. To render a pixel of an image, first sample 3D points along the corresponding ray, and then query the SDF and attenuation coefficient of these points using MLP $\Theta_{sdf}$ and $\Theta_{att}$. 
A surface-bound function was calculated from the SDF and used to constrain attenuation, retaining the attenuation only inside the surface. Then, attenuations are accumulated for volume rendering. 
The final pixel intensity is compared with the ground truth by MSE loss, which is backpropagated to optimize MLPs.}
\label{neas}
\end{center}
\end{figure*}
%

%%%%%%%%%%%%%%%%%%%%%%%%%%%%%%%%%%%%%%%%%%%%
\section{Method}
\label{sec:methodology}
We adopt an implicit neural modeling approach to reconstruct 3D scenes with X-ray images.
Fig.~\ref{neas} provides an overview of our method termed NeAS. 
Given a set of input X-ray images ${\mathcal{I}^{\mathrm{X}}} = \{I_i\}(i=1,2,...,n^{\mathrm{X}})$, our method constructs attenuation fields $\Theta_{\mathrm{att*}}$ and an implicit SDF $\Theta_{\mathrm{sdf}}$. 
$n^{\mathrm{X}}$ is the number of input images. 
The attenuation field determines the attenuation coefficient at each point within the scene, whereas the SDF continuously represents the surface, providing a detailed and continuous geometric description.
The trained SDF offers a 3D geometric constraint to the attenuation field.

%%%%%%%%%%%%%%%%%%%%%%%%%%%%%%%%%%%%%%%%%%%
\subsection{Neural Attenuation Surface}
\label{sec:neas}
\subsubsection{Overview}
Let us begin by considering only one concerning surface: the object--air surface.
Here, we assumed that the attenuation of the signal in air is sufficiently small to be distinguishable from that of the object. 
We assumed that the initial parameters of the intrinsic and extrinsic parameters $(\mathbf{K}, \mathbf{p}_i)$ of the input image $I_i$ are given by calibration. 
The detail of the projection model and the pose refinement are described in Section \ref{sec:pose_refinement}. 
We represented a scene with MLPs: an attenuation field $\Theta_{\mathrm{att}}$ and an SDF $\Theta_{\mathrm{sdf}}$ (Fig.~\ref{neas}).

We sampled a 3D point $\mathbf{x} \in \mathbb{R}^3$ along a ray $\mathbf{r} \in \mathbb{R}^3$ corresponding to a pixel $\mathbf{u} \in \mathbb{R}^2$ in an X-ray frame: $\mathbf{x} = \mathbf{o} + t\mathbf{r}$. 
Here, $t$ is the travel distance from the origin $\mathbf{o}$ to the sampled point. 
We omitted the indices for simplicity. 
The input to $\Theta_{\mathrm{sdf}}$ is the encoded position, $\Gamma (\mathbf{x})$, where $\Gamma()$ is a positional encoding function.
The output of $\Theta_{\mathrm{sdf}}$ is the intermediate $K$-dimensional feature vector $\mathbf{f} \in \mathbb{R}^K$ and signed distance $d \in \mathbb{R}$.
$\mathbf{f}$ is the input of the attenuation field $\Theta_{\mathrm{att}}$.
We applied a surface-bound function $\Omega (d,s)$ to the obtained signed distance with a learnable parameter $s \in \mathbb{R}$ to retain the attenuation only inside the surface.
Finally, the network outputs the attenuation coefficient at $\mathbf{x}$ as $\mu(\mathbf{x})$ by multiplying the bounded signed distance by the attenuation factor $\Bar{\mu}$, which is the output of $\Theta_{\mathrm{att}}$.

The method obtains the pixel intensity corresponding to a ray with all sampled points through volume rendering and compares the intensity with the ground truth. 
The detail of the the rendering model is described in \ref{sec:rendering}. 
We applied a ray-casting approach~\cite{roth1982ray} to calculate the rays’ transmittance as they traveled through the scene during volume rendering.
The backpropagation process optimized $\Theta_{\mathrm{sdf}}$ and $\Theta_{\mathrm{att}}$.
The following sections explain the key functions of this framework.

\begin{figure}[t]
\begin{center}
\includegraphics[width=1.0\linewidth]{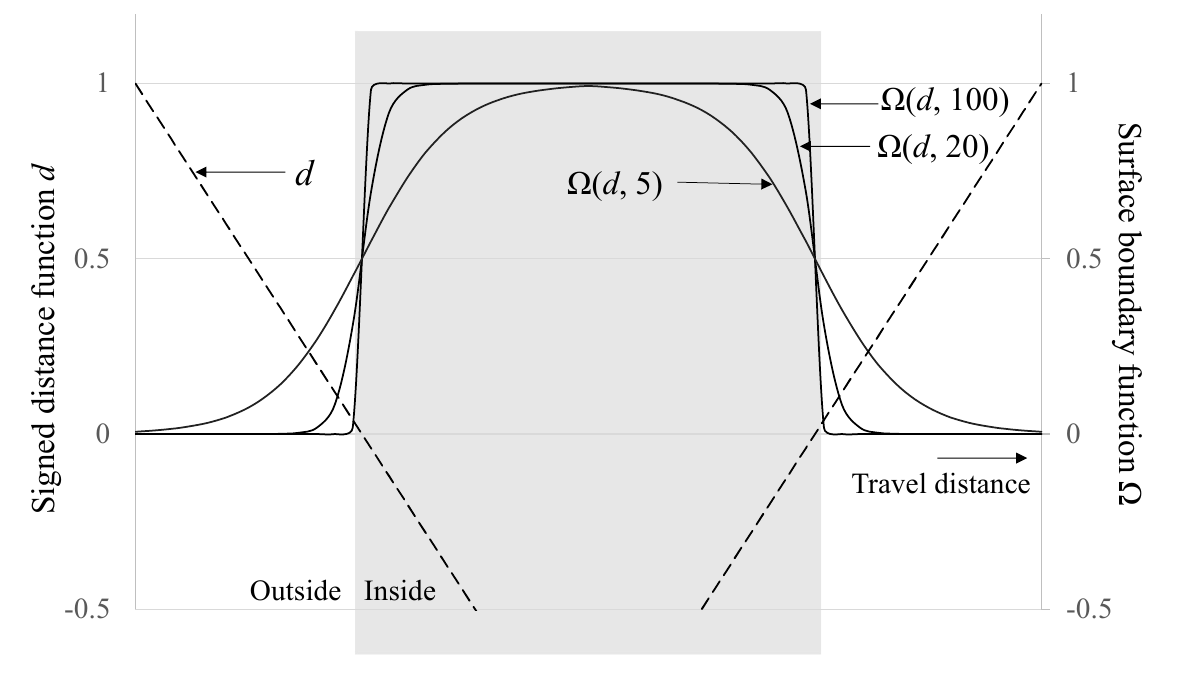}
\caption{Surface boundary function (SBF) 
$\Omega(d,s)$ with SDF values $d$ on a certain ray and different $s$ parameters. As the $s$ value increases, the more accurate the surface position is determined.}
\label{sdf}
\end{center}
\end{figure}
\begin{figure*}[t]
\begin{center}
\includegraphics[width=0.9\linewidth]{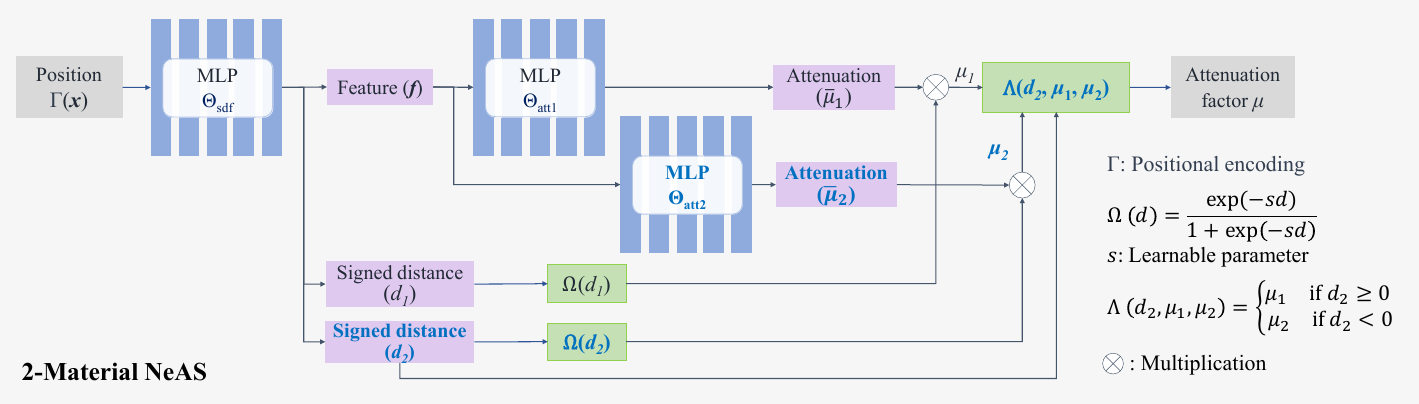}
\label{fig:2sdf}
\end{center}
\caption{Pipeline to query attenuation coefficient for two materials.}
\end{figure*}
%
%%%%%%%%%%%%%%%%%%%%%%%%%%%%%%%%%%%%
\subsubsection{Sampling and encoding}
We employed a stratified method~\cite{nerf} to sample the 3D points along a ray within a fixed range. 
The stratified sampling method divided the points into several mutually exclusive and inclusive ``strata'' and randomly sampled from each stratum.
Note that we did not use a hierarchical sampling strategy because it was not suitable for X-ray conditions, where each point along the ray contributed uniformly, unlike in RGB images, where the points closer to the surface have major contributions.

We applied positional encoding methods, which encoded the input position into a higher-dimensional space to enable the learning of high-frequency variations in intensity and geometry.
The encoding approach addressed the MLP's inclination toward lower frequency functions~\cite{rahaman2019spectral}.
We implemented two encoding techniques: frequency encoding~\cite{nerf} and hash encoding~\cite{instantngp}.

\noindent\textbf{Frequency encoding} involves the use of sine and cosine functions to facilitate high-dimensional mapping, as utilized in the original NeRF.
A mapping function $\gamma$ maps the input value $p \in \mathbf{x}$ from $\mathbb{R}$ to $\mathbb{R}^{2L}$:
\begin{equation}
\gamma(p)\! =\! (\! sin(2^0\pi p),\! cos(2^0\pi p),...,\! sin(2^{L-1}\pi p), \! cos(2^{L-1}\pi p)),
\end{equation}
where $L$ determines the highest frequency.

\noindent\textbf{Hash encoding} leverages a multiresolution voxel grid structure to store feature embeddings with a hash table, which significantly reduces memory requirements for maintaining this lookup table.
The detailed implementation is the same as that of Instant-NGP \cite{instantngp}.

%After encoded, the 3D coordinates of sampled points are fed into two sequential MLPs.
%The first, an SDF MLP, outputs the SDF value for the input point, and its final layer's feature embedding is fed into the subsequent ATT MLP. This second MLP is responsible for determining the attenuation coefficient of each point. 

%%%%%%%%%%%%%%%%%%%%%%%%%%%%%%%%%%%%%%
%\subsubsection{Signed distance function}
\subsubsection{Surface boundary function}
SDF predictions were utilized to establish a surface boundary function (SBF) that could further modify the attenuation coefficient obtained from $\Theta_{\mathrm{att}}$.
As depicted in Fig.~\ref{sdf}, for an arbitrary ray, the ray intersects the object surfaces twice, indicating surface boundaries at these points.
We defined the SBF by resembling a sigmoid function as follows:
\begin{equation}
\label{eq:surface_boundary}
\Omega(d,s) = \frac{exp(-sd)}{1+exp(-sd)},
\end{equation}
where $s$ is a learnable parameter governing the steepness of the surface boundaries.
The variation of the boundary function with parameter $s$ is shown in Fig.~\ref{sdf}.

Ideally, the SBF should be a unit step function equal to 1 inside the object and 0 outside, so that only the attenuation within the object's surface is preserved (see also Eq.~(\ref{eq:att})), effectively disregarding the external attenuation.
However, the step function has challenges in terms of gradient propagation during training.
To address this, we approximated the step function with a sigmoid function, which closely resembled the desired behavior and facilitated smoother gradient propagation, thereby enhancing the model's learning efficiency.

%%%%%%%%%%%%%%%%%%%%%%%%%%%%%%%%%%%%%%%%%
\subsubsection{Neural attenuation field}
By using the vector $\mathbf{f}$ output from $\Theta_{\mathrm{sdf}}$ as the input, $\Theta_{\mathrm{att}}$ outputs the attenuation parameter $\bar{\mu} \in \mathbb{R}$.
By applying the SBF (Eq.~(\ref{eq:surface_boundary})) to the parameter, we obtained the detailed attenuation coefficient $\mu \in \mathbb{R}$ at the sampled point $\mathbf{x}$ as follows:
\begin{equation}
%\ \ \ \ \ \ \ \ \ \mu = \frac{exp(-sd)}{1+exp(-sd)}\mu_d,
\mu(\mathbf{x}) = \Omega(d,s)\bar{\mu}.
\label{eq:att}
\end{equation}
%
%where $\bar{\mu}$ represents the output of the ATT MLP and $\mu$ signifies the final predicted attenuation coefficient for a specific point, which will be used in later volume rendering. 

$\Theta_{\mathrm{att}}$ must produce an output $\bar{\mu}>0$ to retain attenuation only within the interior of the surface.
Therefore, we employed the activation function $\alpha\sigma(x)+\beta$ in the last layer of $\Theta_{\mathrm{att}}$. Here, $\sigma(\cdot)$ denotes the sigmoid function and $\alpha (>0) \in \mathbb{R}$ and $\beta (>0) \in \mathbb{R}$ are parameters that are adjusted based on the attenuation characteristics of the object.
Thus, the range of the attenuation parameter $\bar{\mu}$ was $[\beta,\beta+\alpha]$. $\beta$ determines the minimum attenuation value and ensures that it is not zero.
This constraint is crucial because only the SBF is designed to determine the regions of nonzero attenuation (where the signed distance $d<0$) and areas with zero attenuation (where $d>0$).
If $\beta$ were set to zero, the final attenuation coefficient $\mu(\mathbf{x})$ could be $0$ when $\bar{\mu}=0$ even in regions where $d<0$.
In such cases, the SBF cannot represent accurate boundaries.
%Therefore, it is necessary for $\Theta_{\mathrm{att}}$ to produce an output $\bar{\mu}>0$, then this output is subsequently refined by the surface boundary function, which clips the values to retain attenuation only within the interior of the surface.

%%%%%%%%%%%%%%%%%%%%%%%%%%%%%%%%%%%%%
\subsubsection{Rendering and optimization}
\label{sec:rendering}
Once the attenuation coefficients for all points along a ray were determined, we calculated the pixel intensity corresponding to the ray using volume rendering.
Since X-rays are generally absorbed without being scattered by most materials, Lambert-Beer's law is assumed to hold. 

If the attenuation coefficient $\mu$ is a continuous function of the distance $t$, the observed light intensity $\hat{I}(\mathbf{r})$ after traveling a distance $T$ can be written as follows: 
%The intensity for a specific ray that has traversed the target object is computed using the following equation:
%
\begin{equation}
\hat{I}(\mathbf{r}) = I_0(\mathbf{r}) \int_{0}^{T} exp(-\mu(t))dt,
\label{xray_equa}
\end{equation}
where $I_0(\mathbf{r})$ is the light intensity emitted from the X-ray source to the ray direction $\mathbf{-r}$.

%where $\mu(t)$ represents the linear attenuation coefficient at distance $t$, $I_0$ is the intensity emitted from the X-ray source, and $I$ is the X-ray's final intensity after traveling a distance $r$. 
%We can assume that $I_0 = 1$ because image intensities are normalized to $[0,1]$. 
Assuming that the attenuation in the air is sufficiently small, $I_0$ can be regarded as the maximum pixel value: $I_0 = 1$. 
By applying the quadrature rule \cite{optical}, we can approximate the calculation of the integral while maintaining differentiability for optimization purposes:
\begin{equation}
\hat{I}(\mathbf{r}) = exp(-\sum_{i=j}^{N}\mu(\mathbf{x}_j)\delta_j),
\label{dis_xray}
\end{equation}
where $\delta_j$ is the adjacent distance between $t_{j+1}$ and $t_j$ and $N$ is the total number of sampled points on the ray $\mathbf{r}$.

%$t_{i+1}$ and $t_i$ are the ray depth for point $i+1$ and $i$, $\mu_i$ is the attenuation at point $i$,

Finally, the predicted ray intensity $\hat{I}(\mathbf{r})$ is compared with the input pixel intensity by calculating the MSE loss:
\begin{equation}
\begin{aligned}
%\mathcal{L}_{\mathrm{int}} = \sum_{\mathbf{r}\in\mathcal{R}}\Big[\Big\Vert \hat{I}(\mathbf{r})-I(\mathbf{r})\Big\Vert_2^2\Big], 
\mathcal{L}_{\mathrm{int}} = \sum_{\mathbf{r}\in\mathcal{R}}\Big\Vert \hat{I}(\mathbf{r})-I(\mathbf{r})\Big\Vert_2^2, 
\end{aligned}
\end{equation}
where $\mathcal{R}$ is a sampled batch of the rays.
We applied an Eikonal term \cite{eikonal} to regularize SDF as follows:
% \color{red}please define $m$, $n$, $\mathbf{n}$\color{black}
%
\begin{equation}
\begin{aligned}
\mathcal{L}_{\mathrm{reg}} = \frac{1}{mn}\sum_{k,j}(\Vert \mathbf{n}_{k,j}\Vert - 1)^2.
\end{aligned}
\end{equation}
where $m$ is the batch size, $n$ is the point sampling size, and $\mathbf{n}$ is the gradient of the SDF field at the sampling point.
Then, the final loss for the optimization is calculated as follows:
\begin{equation}
\begin{aligned}
\mathcal{L} = \mathcal{L}_{\mathrm{int}} + \lambda\mathcal{L}_{\mathrm{reg}},
\end{aligned}
\end{equation}
where $\lambda$ is a weighting parameter.
\begin{figure}[t]
	\begin{center}
		\includegraphics[width=0.75\linewidth]{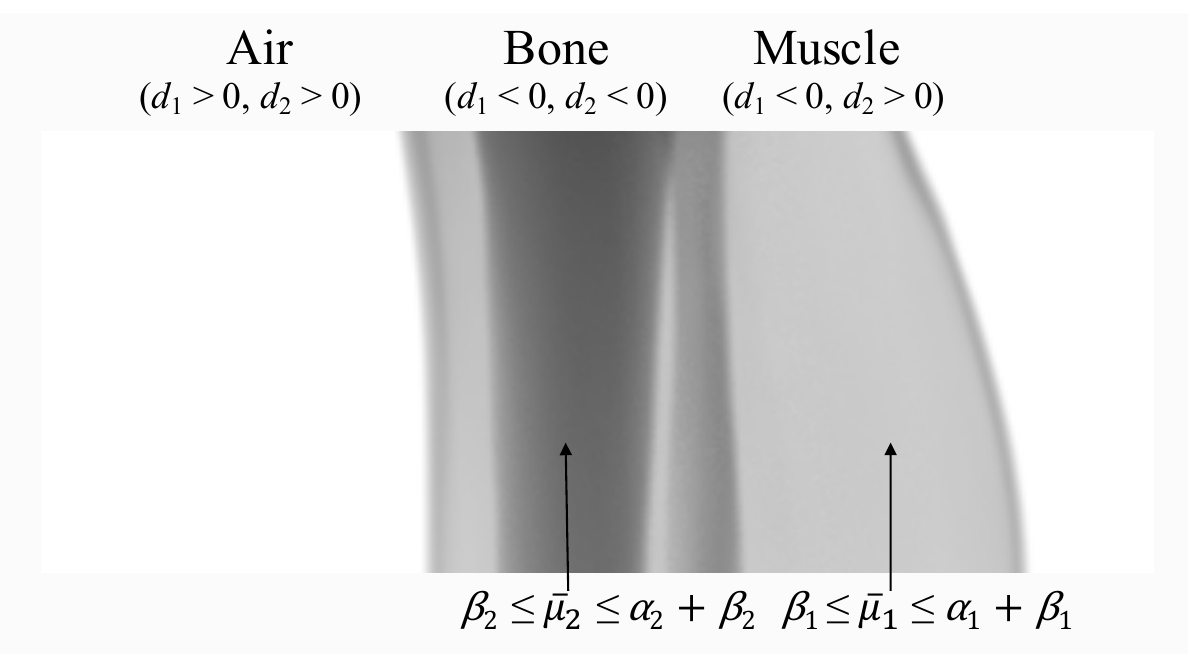}
			\caption{Brief illustration of the attenuation coefficient in a bone and muscle area.}
		\label{muscle_bone}
	\end{center}
\end{figure}
%
%
%%%%%%%%%%%%%%%%%%%%%%%%%%%%%%%%%%%%%%%
\subsection{NeAS for Multiple Materials}
In the case of the human body, the boundaries between multiple materials must be distinguished.
For example, as illustrated in Fig.~\ref{muscle_bone}, an X-ray image of the leg, in addition to the outer skin--air surface, shows that the internal bone--muscle surface is also prominently visible. 
%, although it resides within the skin--air surface.
The approach outlined in the previous section was designed to extract only the surfaces of one material. 
%between two materials: air--skin or muscle--bone surfaces. 
%while omitting the inner bone surfaces.
Therefore, we proposed NeAS for two materials (2M-NeAS) with different attenuation coefficients.

%%%%%%%%%%%%%%%%%%%%%%%%%%%%%%%%%%%%%
\subsubsection{Framework}
2M-NeAS represents two signed distance functions using one MLP $\Theta_{\mathrm{sdf}}$, and the attenuation fields using independent MLPs $\Theta_{\mathrm{att1}}$ and $\Theta_{\mathrm{att2}}$ for each field.
Fig.~\ref{fig:2sdf} shows the pipeline of the proposed method.
$\Theta_\mathrm{sdf}$ models two SDFs and outputs the signed distances for both materials in addition to the intermediate feature vectors $d_1 \in \mathbb{R}$, $d_2 \in \mathbb{R}$ and $\mathbf{f} \in \mathbb{R}^K$.
For example, $d_1$ represents the skin--air surface, whereas $d_2$ represents the bone--muscle surface.

We assigned two separate MLPs for the two attenuation fields.
If we used only one MLP for the attenuation fields, a point near the boundary of two materials would be influenced by both fields.
Modifications to the SDF during training would cause the boundary to shift continuously.
Consequently, the output of the MLP at these points would exhibit significant fluctuations.
Therefore, we added an MLP to represent another attenuation field.
%
%to determine another attenuation inside and outside the surface.
%
%Figure~\ref{fig:2sdf} illustrates the process to predict the attenuation coefficient of a sampled point $\mathbf{x}$ in the presence of two surfaces. 
%
%The distances contribute to surface boundary estimation. 
%The final layer feature of $\Theta_{\mathrm{sdf}}$ is then fed into two distinct MLPs,
$\Theta_{\mathrm{att1}}$ and $\Theta_{\mathrm{att2}}$ are attenuation fields to predict the attenuation parameters $\bar{\mu}_1$ and $\bar{\mu}_2$, respectively.
%$\Theta_{\mathrm{att1}}$ and $\Theta_{\mathrm{att2}}$take $\mathbf{f}$ as input and output attenuation parameters $\hat{\mu}_1$ and $\hat{\mu}_2$, respectively. 
%Two different attenuation fields are learnt.
We applied Eq.~(\ref{eq:att}) and then obtained the attenuation coefficients as $\mu_1=\Omega(d_1,s)\bar{\mu}_1$ and $\mu_2=\Omega(d_2,s)\bar{\mu}_2$.

The activation function for $\Theta_{\mathrm{att1}}$ and $\Theta_{\mathrm{att2}}$ is $\alpha\sigma(x)+\beta$. Thus, the value ranges of $\bar{\mu}_1$ and $\bar{\mu}_2$ differ: $(\beta_1,\alpha_1+\beta_1)$ and $(\beta_2,\alpha_2+\beta_2)$.
Tailoring these parameters to the specific characteristics of the materials being imaged is crucial to ensure that the ranges of muscle and bone attenuation do not overlap.
Nonoverlapping ranges are essential for maintaining the distinctiveness of SDFs in accurately representing respective surfaces.

\subsubsection{Selecting function}

Because the proposed framework generates two attenuation coefficients $\mu_1$ and $\mu_2$, we must select the proper one for the point $\mathbf{x}$.
%Subsequently, 
Therefore, we introduce a selector mechanism to determine which of the two attenuation values may be chosen as the final prediction as follows:
%
% \begin{equation}
% \mu=
% \begin{cases}
% \frac{exp(-sd_2)}{1+exp(-sd_2)}\mu_{d2}& \text{$d_2<0$}\\
% \frac{exp(-sd_1)}{1+exp(-sd_1)}\mu_{d1}& \text{else}.
% \end{cases}
% \end{equation}
%
\begin{equation}
\mu=\Lambda(d_2,\mu_1,\mu_2)=\begin{cases}
\mu_1 & (\text{$d_2 \geq 0$})\\
\mu_2 & (\text{$d_2 < 0$}).
\end{cases}
\end{equation}
This approach implies the following: if this point lies inside the bone where $d_2<0$, we output $\mu_2$ as the final attenuation; if this point belongs to the muscle, $\mu_1$ is output instead.
This distinction is dynamic and may evolve as the SDFs are optimized during the training process.
Fig.~\ref{muscle_bone} presents a conceptual visualization of the leg-part X-ray image.
The bone is surrounded by muscle, and $d_2<0$, the final attenuation should be $\mu_2$.
Outside the bone, where $d_2 \geq 0$, everything is the same as in Section~\ref{sec:neas}.

%%%%%%%%%%%%%%%%%%%%%%%%%%%%%%%%%%%
\subsection{Pose Refinement}
\label{sec:pose_refinement}
%To render a pixel in a new view image, we firstly compute the direction and origin of its corresponding ray $\mathbf{r}$. 
% For the establishment of an exact implicit field and the rendering of precise X-ray images, accurate camera poses is essential.
% However, even after calibration, there may be residual errors in camera poses. 
% To rectify this, a pose refinement technique is employed to further fine-tune these poses. 

Because accurate projection parameters are essential for traversing the ray, we optimized and refined the poses of the X-ray source and flat panel detector (FPD).
In an X-ray imaging system using robotic arms or a turntable, even after calibration, we cannot avoid residual errors in the poses of the X-ray source and FPD relative to the target.
Therefore, we refined the poses for all input X-ray images simultaneously while training the NeAS.
The entire rendering process was differentiable, which allowed the loss to backpropagate and adjust the camera poses, enabling concurrent optimization along with network parameters.
%, a process detailed in a subsequent section.

\subsubsection{Camera model}
The X-ray system can be regarded as a pinhole camera system, where the X-ray source functions as the optical center, and the FPD serves as the image plane.
The measurement system did not include lenses; therefore, we did not consider lens distortion.
%Considering that rotation is controlled by computer precisely, 
Our focus in pose refinement was primarily on the uncertainties in the object--source distance and source--panel relative position.
Consequently, we optimized only the translation components in the extrinsic parameters and principal points in the intrinsic parameters.
%\color{red}please explain what parameters do you refine. \color{black}
Here, we assumed that reasonable initial parameters are given.

\subsubsection{Coarse-to-fine strategy}
%In the case of usual photo, only the surface of the object is visible and opaque with continuous surface color.
In X-ray imaging, not only the surface point contributes to the final pixel value, and the inner points along the ray also play important roles.
That is, even minor changes in the viewing angle can lead to substantial changes in the rendering results, as all points along a given ray can shift drastically.
In other words, refining poses using a gradient-based approach by back-propagation is challenging. 

To address this problem, we incorporated a frequency regularization approach by integrating a weight mask into the encoding process, as has been done previously \cite{neus2,yang2023freenerf,li2023neuralangelo}. 
Frequency regularization has also been proven to be effective in grid-based feature encoding~\cite{xie2024g}. 
At the beginning of the training, we were only concerned with low-frequency outlines, and the rendering results were predominantly derived from the lower-frequency part of the input.
As training progressed, high-frequency features were utilized.
%We can distinguish such low-frequency and high-frequency components from positional encoding. 
The mask reweighted the components in the $k$th frequency as follows:
\begin{equation}
%\gamma_{reweight_k}(x) = w_k(\tau)\gamma_k(x),
\hat{\Gamma}_k(p) = w_k(\tau)\Gamma_k(p),
\end{equation}
where $\Gamma_k$ can be the frequency encoding at frequency $k$ or hash encoding at the $k$th resolution.
$w_k$ is defined as
\begin{equation}
w_k(\tau)=
\begin{cases}
0& \tau<k\\
\frac{1-cos((\tau - k)\pi)}{2}& 0\leq\tau - k<1 \\
1& \tau - k\geq1,
\end{cases}
\end{equation}
where $\tau$ is a parameter that increases in relation to the number of iterations during training.
We also took a worming-up strategy; the pose refinement started after $M$ (=500 in the experiment) iterations to avoid the early overfitting problem.
%This coarse-to-fine mask functions like a kind of low-pass filter, when $\tau$ is lower than frequency $k$, $w_k(\tau)=0$, effectively suppressing high-frequency components.
%This approach allows pose refinement to initially concentrate on the low-frequency general outline of the image and subsequently shift focus to high-frequency details.
%
% \begin{figure}[t]
% \captionsetup{justification=centering}
% 	\begin{center}
% 		\includegraphics[width=0.8\linewidth]{graph/xray.png}
% 			\caption[labelInTOC]{X-ray image simulation by gVirtualXray \cite{gvxr}.}
% 		\label{fig:xray}
% 	\end{center}
% \end{figure}
% %
%
\begin{figure}[t] 
\begin{center}
  \includegraphics[width=1\linewidth]{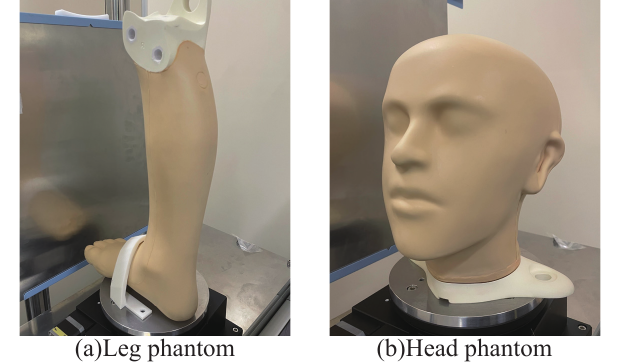}  
\end{center}
    \caption{Phantoms used for X-ray image capturing.}
  \label{phantom}
\end{figure}

\begin{figure}[t]
    \begin{center}
    \includegraphics[width=1\linewidth]{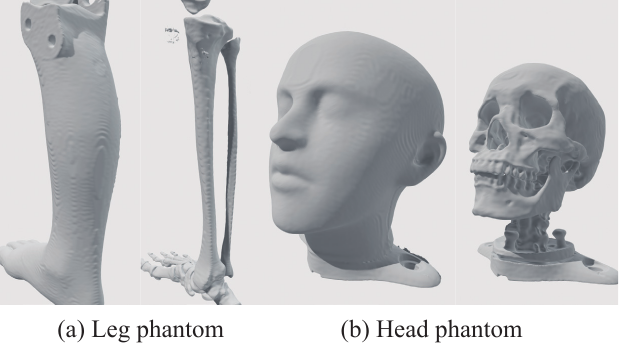}        
    \end{center}
    \caption{3D meshes of phantoms skin and bone extracted by Helical CT system}
    \label{fig:ct_refmesh}
\end{figure}

\section{Experimental Results}
%\subsection{Experimental Settings}
%\textbf{Data.} 
\subsection{Dataset}
We conducted experiments using simulated and real X-ray images.
We employed gVirtualXray \cite{gvxr} to obtain the simulated X-ray images.
%Figure~\ref{fig:xray} shows an example of a captured image of a knee bone from a given viewpoint.
We generated 36 images of size $480\times480$ around the 3D models of the knee bone~\footnote{embodi3d. (2017). Right knee - Bone model [3D model]. Retrieved from https://www.embodi3d.com/files/file/12503-right-knee-bone-model-stl-file-from-converted-ct-scan/ (Licensed under CC)} and skull~\footnote{jagake. (2022). JAGAKE Full Skull [3D model]. Sketchfab. Retrieved from https://sketchfab.com/3d-models/jagake-full-skull-a69cb39385ae4cbbad7962a816fd8c06 (Licensed under CC Attribution)} derived from CT scans, altering the view direction by 10$^{\circ}$.
The trajectory of the X-ray source position is a circle on the $x$--$y$ plane.
From a total of 36 images, five were selected for validation purposes, and the model was trained using the remaining 31 images.
The distance from the source to the object's center was $40 ~\text{cm}$, whereas that from the center of the object to the detector panel was $10~\text{cm}$.

We captured 30 real images around a phantom using a computer-controlled turntable panel, with a 12$^{\circ}$ change in the angle for each image. Phantoms include a leg and a head phantom as shown in Fig.~\ref{phantom}. Fig.~\ref{fig:ct_refmesh} shows the 3D meshes of the phantoms bone and skin obtained by Helical CT scan using Aquilion Prime SP provided by Canon Medical Systems.
Out of 30 views, four images were designated as validation images, whereas the remainder were used for training.
The distance between the source and FPD was approximately $1~\text{m}$, whereas the distance from the X-ray source to the object was approximately $77~\text{cm}$.
Note that these distances were not exact, resulting in inaccuracies in the camera parameters.
Pose refinement was used to compensate for these inaccuracies.
The X-ray source voltage was set at 60 kV, and the current was set at 200 mA, ensuring that the images were relatively clear and minimally affected by noise.

\begin{figure*}[t]
\begin{center}
\includegraphics[width=1.0\linewidth]{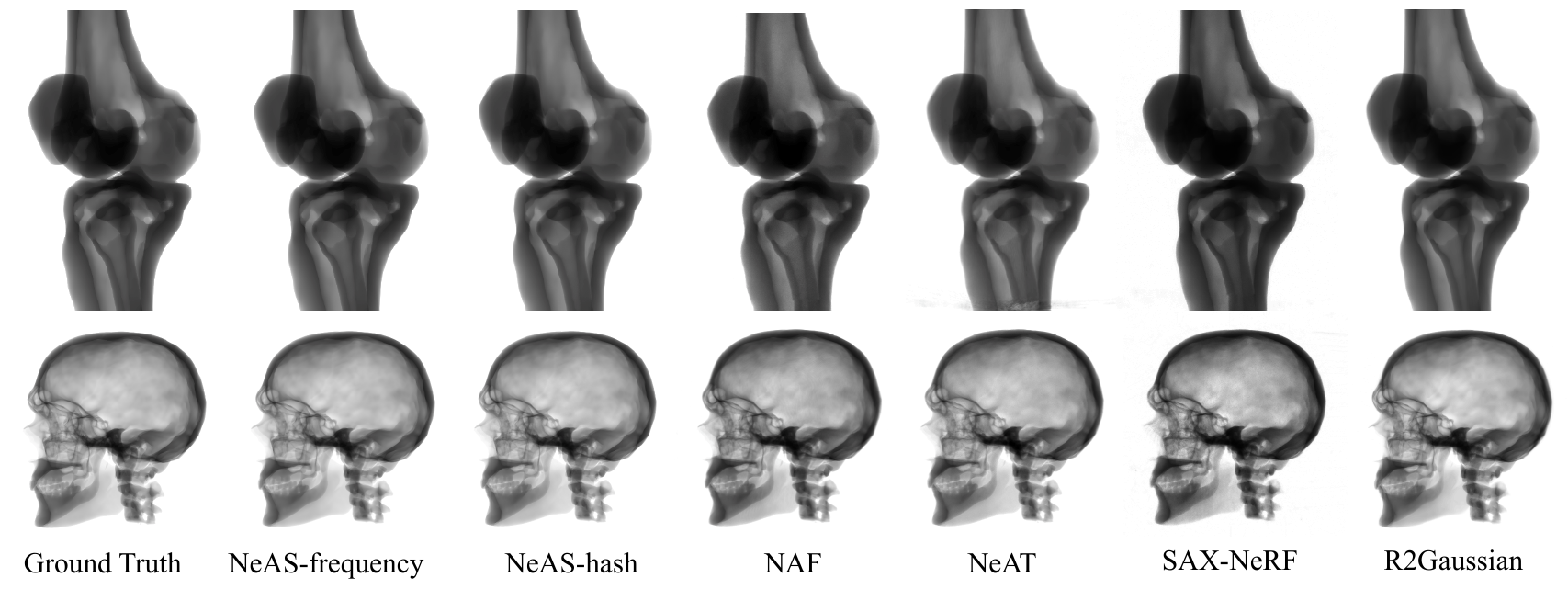}
\label{fig:knee}
\end{center}
\caption{Validation views for simulation dataset. }
\end{figure*}
%
%
% \begin{figure}[t]
% \captionsetup{justification=centering}
% 	\begin{center}
% 		\includegraphics[width=.8\linewidth]{graph/skull_2502.pdf}
% 			\caption[labelInTOC]{2D validation views for simulation data on a skull}
% 		\label{skull}
% 	\end{center}
% \end{figure}
% %
%

\subsection{Implementation Details}
We assumed that the region of interest lies within a unit sphere; accordingly, the scene was initially scaled to fit within this sphere.
Image pixel values were normalized to the range of $[0,1]$, with the background set to $1$ where the rays did not undergo attenuation. 
In each batch, 512 rays were randomly sampled and 128 points were sampled along each ray. 
For frequency encoding, $\Theta_{\mathrm{sdf}}$ comprised six layers, each with a hidden size of 256, whereas $\Theta_{\mathrm{att}}$ included three hidden layers, with a size of 256.
For hash encoding, both $\Theta_{\mathrm{sdf}}$ and $\Theta_{\mathrm{att}}$ were constructed with two layers, each with a hidden size of 64.
The resolution of hash encoding varied across the 14 levels, ranging from 16 to 2048.

For pose refinement, the parameter $\tau$ was first set to 2, so that the first two low-frequency components were of concern at the beginning.
It then grew linearly to the largest frequency $L$ until half of the total number of training iterations.
Parameter $s$ for the SBF was initialized as $20$.
For the 2M-NeAS experiment, we chose $\beta_1 = 0.1$, $\alpha_1 = 3.4$, $\beta_2 = 3.5$, and $\alpha_2 = 5.5$.
The experiments were conducted using a single NVIDIA RTX3090 GPU.

\subsubsection*{Determining $\alpha$ and $\beta$}
The parameters $\alpha$ and $\beta$ are related to the attenuation coefficients of the target objects.
If we can approximate $\mu_{\text{air}}$, $\mu_{\text{muscle}}$, and $\mu_{\text{bone}}$, we can set two thresholds as follows:
\begin{align}    
t_1 &= (\mu_{\text{air}} + \mu_{\text{muscle}})/2 \\
t_2 &= (\mu_{\text{muscle}} + \mu_{\text{bone}})/2
\end{align}
For the muscle, we can set the attenuation range as $[t_1, t_2]$ as follows:
\begin{align}    
\beta_1 &= t_1 \\
%\beta_1 &= (\mu_{\text{air}} + \mu_{\text{muscle}})/2 \\
\alpha_1 &= t_2 - t_1.
%\alpha_1 &= (\mu_{\text{bone}} - \mu_{\text{air}})/2.
\end{align}
For the bone, the attenuation range could be $[t_2, t_{\text{max}}]$, and $t_{\text{max}}$ must be 
larger than the attenuation coefficients of all bone regions:
\begin{align} 
\beta_2 &= t_2 \\
\alpha_2 &= t_{\text{max}} - t_2.
\end{align}

Another method is to run a 1M-NeAS first and set a relatively large range for the attenuation coefficients.
By analyzing the attenuation distribution, we can easily find the threshold values.
%That's the way that how we decided them in my experiments.
%
%
\begin{figure*}[t]
\begin{center}
\includegraphics[width=1.0\linewidth]{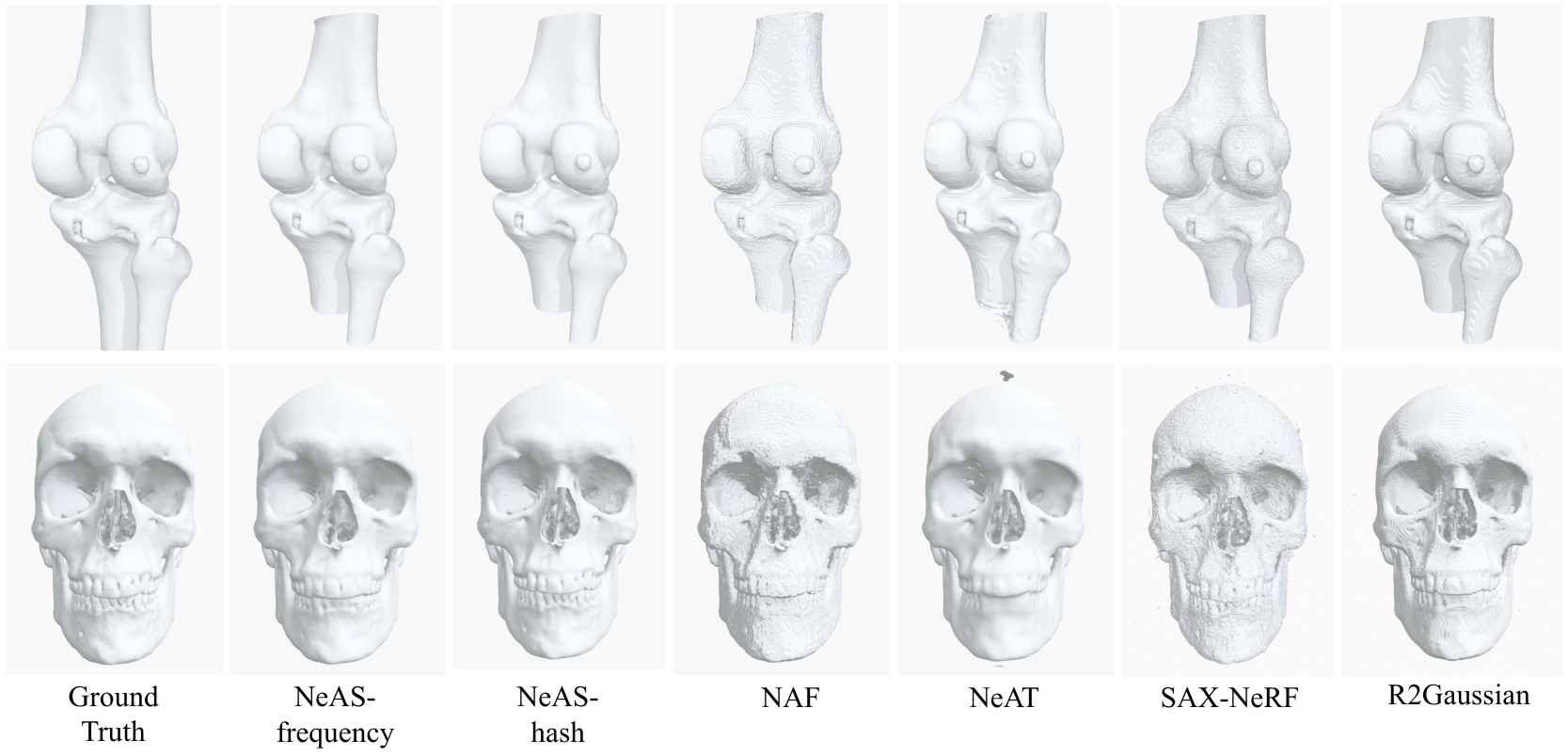}
    %,
%as is shown in Fig.~\ref{fig:xray}, 
%resulting in the truncation of the reconstructed results at the upper and lower ends. 

\label{fig:knee_mesh}
\end{center}
    \caption[labelInTOC]{Extracted knee-bone and skull surfaces. Note that reconstructed knee-bone models are partly included because of cone-beam imaging.}
\end{figure*}

%			NeAS frequency & \textbf{\textcolor{red}{53.20}} & \textbf{\textcolor{red}{0.9996}} & \textbf{\textcolor{red}{0.0017}} & \textcolor{blue}{0.1877} \\
%			NeAS hash & \textcolor{blue}{52.71} & \textcolor{blue}{0.9995} & \textcolor{blue}{0.0027} & \textbf{\textcolor{red}{0.1462}} \\
%
\begin{table}[tb]
	\begin{center}
	\caption{Quantitative comparison on knee simulation data (average of five validation views)}
	\label{tab:knee_psnr}
		\begin{tabular}{l|cccc}\hline
			& PSNR  $\uparrow$ & SSIM $\uparrow$ & LPIPS $\downarrow$ & CD (mm) $\downarrow$ \\
            \hline
			NeAS frequency & \textcolor{red}{\textbf{53.20}} & \textcolor{red}{\textbf{0.9996}} & \textcolor{red}{\textbf{0.0017}} & \textcolor{blue}{0.1877} \\
            NeAS hash & \textcolor{blue}{52.71} & \textcolor{blue}{0.9995} & \textcolor{blue}{0.0027} & \textcolor{red}{\textbf{0.1462}} \\
			NAF \cite{naf} & 44.49 & 0.9974 & 0.0441 & 0.6082 \\
			NeAT \cite{neat} & 48.37 & 0.9954 & 0.0424 & 0.2244 \\
            SAX-NeRF \cite{sax_nerf} & 42.56 & 0.9824 & 0.0064 & 0.2432 \\
            R${}^2$-Gaussian \cite{r2_gaussian} & 47.12 &0.9970 & 0.0095 & 0.4021 \\
            \hline
        \end{tabular}
	\end{center}
\end{table}

\begin{table}[t]
	\begin{center}
	\caption{Quantitative comparison on skull simulation data (average of five validation views)}
	\label{tab:skull_psnr}
    \begin{tabular}{l|cccc}\hline
			& PSNR $\uparrow$ & SSIM $\uparrow$ & LPIPS $\downarrow$ & CD (mm) $\downarrow$ \\ \hline
                NeAS frequency & \textcolor{blue}{48.35} & \textcolor{blue}{0.9985} & \textcolor{red}{\textbf{0.0063}} & \textcolor{blue}{0.3521} \\
			NeAS hash & \textcolor{red}{\textbf{49.29}} & \textcolor{red}{\textbf{0.9986}} & \textcolor{blue}{0.0186} & \textcolor{red}{\textbf{0.167}} \\
			NAF \cite{naf} & 40.43 & 0.9877 & 0.0993 & 0.7390 \\
			NeAT \cite{neat} & 46.59 & 0.9931 & 0.0655 & 0.3640 \\
            SAX-NeRF \cite{sax_nerf} & 32.34 & 0.9082 & 0.1646 & 0.2871 \\
            R${}^2$-Gaussian \cite{r2_gaussian} & 32.80 & 0.9333 & 0.1415 & 0.3855 \\
            \hline
        \end{tabular}
	\end{center}
\end{table}
            
\subsection{Results on Simulation Data}
We first evaluated the performance of our NeAS method on simulation data.
There was only one surface of concern, so we utilized 1M-NeAS for this experiment.
We used known camera poses to render the simulated images.
We tested our NeAS in the knee-bone scene with frequency encoding and hash encoding.

\begin{figure*}[t]
    \centering
    \includegraphics[width=0.95\linewidth]{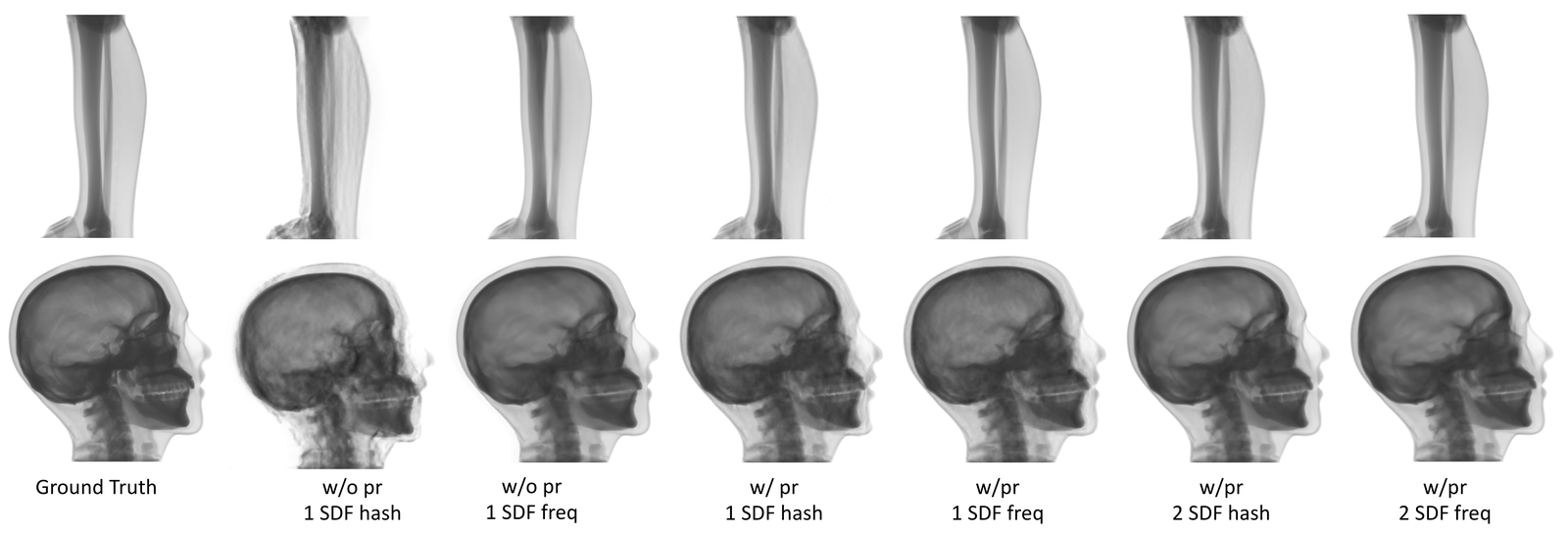}
    \caption[labelInTOC]{2D validation views for real leg and head X-ray image data. The results are compared with and without pose refinement, between hash encoding and frequency encoding, and between a 1 SDF setting and 2 SDF setting.}
    \label{real_data}
\end{figure*}

%%%%%%%%%%%%%%%%%%%%%%%%%%%%%%%%%%%%%%%%%%%%
\subsubsection{Novel view synthesis}
Figs.~\ref{fig:knee} shows the synthesis results for validation views of knee and skull bones, alongside comparisons with NAF \cite{naf}, NeAT \cite{neat}, SAX-NeRF~\cite{sax_nerf}, and R${}^2$-gaussian~\cite{r2_gaussian}.
The 2D results from NAF and NeAT appeared to fail to 
similarly have high quality at first glance. 
However, upon closer inspection, the results of NAF, NeAT, and SAX-NeRF seem to emphasize darker regions, indicating that the attenuation is estimated to be higher than correct values. 
The results of the SAX-NeRF show the artifacts in the area other than the object. 

A detailed quantitative comparison is shown in Tables \ref{tab:knee_psnr} and \ref{tab:skull_psnr}. 
The values are the average of five validation views.
In the tables, Peak Signal-to-Noise Ratio (PSNR), Structural Similarity Index Measure (SSIM), and Learned Perceptual Image Patch Similarity (LPIPS) are metrics used to evaluate image similarities with the ground truth. 
The results revealed that NeAS provided superior performance in comparison to the other methods.

Meanwhile, the results of frequency encoding and hash encoding showed that their performances were nearly the same.
However, the training speed for hash encoding was about 3.5 times that of frequency encoding.

%%%%%%%%%%%%%%%%%%%%%%%%%%%%%%%%%%%%%%%%%%
\subsubsection{Surface reconstruction}
\noindent \textbf{Qualitative evaluation: }
Figs.~\ref{fig:knee_mesh} shows the results of the surface extraction of knee and skull bones, achieved by locating the zero-level set of SDF. 
%using our method, 
The figures also show the surfaces extracted by NAF, NeAT, SAX-NeRF, and R${}^2$-Gaussian for comparison. 
We extracted the surface models of NAF, NeAT, and SAX-NeRF by manually setting the optimal density values and applying the Marching Cubes algorithm. 
The results prove that our method successfully reconstructed the surfaces accurately and smoothly.

%, and NeuS. 
%NeuS was unable to accurately learn the correct surface because of its inability to interpret X-ray images.
Meanwhile, noticeable artifacts were evident on the surface of the NAF, suggesting that this method has difficulty reconstructing 3D geometry and surface, although it performs well in novel view synthesis.
In comparison, although NeAT's result appears reasonable, an artifact was discovered at the top of the skull, and the reconstruction of the teeth still requires improvement.

\noindent \textbf{Quantitative evaluation: }
We also evaluated the results quantitatively using the Chamfer distance (CD) by comparing them with the ground truth. 
As indicated in Tables \ref{tab:knee_psnr} and \ref{tab:skull_psnr}, our method outperformed the other approaches.
Although NeAS with frequency encoding achieved better 2D results, hash encoding outperformed significantly in terms of the CD.
We found that the surface extracted by the NeAS frequency was over-smoothed, and that some details were ignored.
This may be attributed to the inherent bias of both $\Theta_{\mathrm{sdf}}$ and $\Theta_{\mathrm{att}}$ toward low-frequency features.
The impact of this issue was less obvious in the knee-bone results, as they lacked significant high-frequency details, whereas the skull contained many high-frequency details.
Conversely, the implementation of hash encoding significantly alleviated this problem.
The multiresolution structure of hash encoding played an extremely important role, with higher-resolution voxel grids focusing on high-frequency details.

\begin{figure*}[t]
\centering
\includegraphics[width=0.85\linewidth]{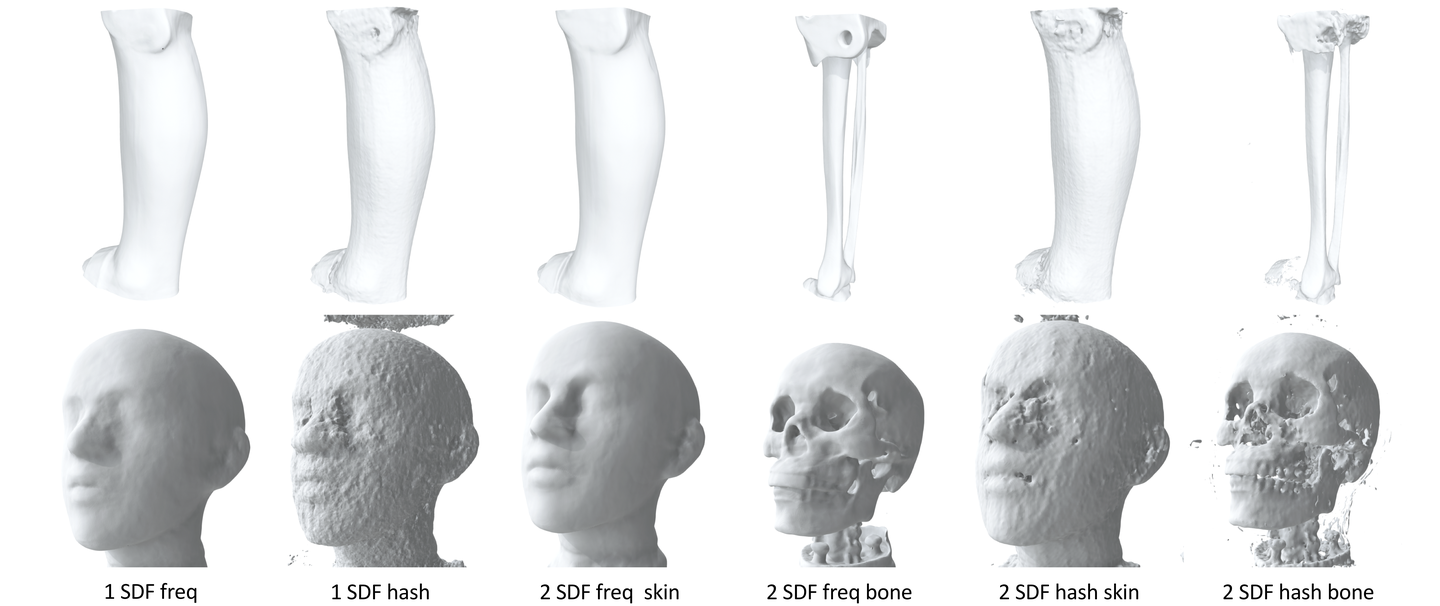}
\caption[labelInTOC]{3D surfaces extracted from real X-ray data. Both 1 SDF and 2 SDF models were implemented. All the results were obtained from refined poses.}
\label{fig:real_surface}
\end{figure*}

\begin{table*}[!ht]
    \centering
    \caption{Quantitative results on a real X-ray dataset.}
    \begin{tabular}{c|c||cc|cc|cc}
    \hline
        \multirow{2}*{Scene} & \multirow{2}*{Evaluation} & w/o pr  & w/o pr  & w/ pr  & w/ pr  & w/ pr  & w/ pr  \\ 
        ~ & ~                      & 1 SDF hash & 1 SDF freq & 1 SDF hash & 1 SDF freq & 2 SDF hash & 2 SDF freq \\ \hline
        \multirow{3}*{Leg} & PSNR$\uparrow$  & 27.19  & 31.35  & 35.89  & \textcolor{red}{\textbf{42.80}}  & 36.04 & \textcolor{blue}{42.40} \\ 
        ~ & SSIM$\uparrow$                   & 0.9325 & 0.9715 & 0.9861 & \textcolor{red}{\textbf{0.9977}} & 0.9854 & \textcolor{blue}{0.9972} \\ 
        ~ & LPIPS$\downarrow $                 & 0.1629 & 0.0701 & 0.0555 & \textcolor{red}{\textbf{0.0216}} & 0.0624 & \textcolor{blue}{0.0236} \\ \hline
        \multirow{3}*{Head} & PSNR$\uparrow$ & 26.69 & 33.81 & 32.30  & \textcolor{red}{\textbf{37.05}}  & 32.72 & \textcolor{blue}{36.38} \\ 
        ~ & SSIM$\uparrow$                  & 0.9139 & 0.9763 & 0.9674 & \textcolor{red}{\textbf{0.9880}} & 0.9681 & \textcolor{blue}{0.9853} \\ 
        ~ & LPIPS$\downarrow $                 & 0.2475 & 0.1316 & 0.1681 & \textcolor{red}{\textbf{0.0978}} & 0.1770 & \textcolor{blue}{0.1067} \\ \hline
    \end{tabular}
	\label{real_quant}
\end{table*}

%\begin{table*}[!ht]
\begin{table}[tb]
    \centering
    \caption{Chamfer distance (\si{mm}) from point cloud extracted from CT on a real X-ray dataset. }
	\label{tab:real_cd}
    \begin{tabular}{c||cc|cc}
    \hline
        Scene & 1 SDF hash  & 1 SDF freq & 2 SDF hash  & 2 SDF freq  \\ \hline
        Skin$\downarrow$  & 2.36  & \textcolor{red}{\textbf{0.34}}  & 2.73  & 0.37 \\  
        Bone$\downarrow$ & - & - & \textcolor{red}{\textbf{0.42}} & 0.67 \\ \hline
        Skin$\downarrow$ & 1.46 & 0.69 & 3.76 & \textcolor{red}{\textbf{0.56}}  \\  
        Bone$\downarrow$ & - & - & \textcolor{red}{\textbf{1.29}} & 1.53 \\ \hline
        
    \end{tabular}
\end{table}

\subsection{Results on Real Data}
We performed experiments on real data taken from leg and head phantoms with 1M-NeAS and 2M-NeAS. 
Because precise camera parameters were not available, we turned to pose refinement.
We first implemented pose refinement with all images including validation images.
Then, fixing the refined camera parameters, we excluded validation images and trained the model solely on training images from the start.
In this way, we can evaluate the results of the validation images quantitatively.

\subsubsection{Novel view synthesis}
Fig.~\ref{real_data} shows the validation images, and Table~\ref{real_quant} presents the quantitative evaluation results.
A comparison of the results with and without pose refinement indicated that pose refinement significantly enhanced the results qualitatively and quantitatively.
This improvement was particularly notable in hash encoding experiments, where imprecise camera parameters could severely degrade the results.
The comparison between 1M-NeAS and 2M-NeAS revealed no significant performance drop with the use of two SDFs and attenuation fields.

\subsubsection{Surface reconstruction}
\noindent \textbf{Qualitative evaluation: }
In the meantime, we could extract the inner surface by learning two SDFs, as shown in Fig.~\ref{fig:real_surface}.
Here, we only show the results with pose refinement.
Qualitative observations indicated that hash encoding faced difficulties with real data.
The reconstructed surfaces appeared rough, and numerous nonexistent points were likely due to hash collisions or noise.
In contrast, frequency encoding was more successful in learning smooth and accurate surfaces, producing results that closely resembled actual objects.
However, the tendency of frequency encoding to generate relatively over-smoothed surfaces persisted.
This was evident in the missing teeth gaps, whereas hash encoding was better in this regard. 

Our observations highlight the need for a balanced approach to encoding techniques, particularly when dealing with intricate details in 3D reconstructions from X-ray images.

\noindent \textbf{Quantitative evaluation: }
Table~\ref{tab:real_cd} shows CD from point clouds extracted from volume data obtained by Helical CT scan using Aquilion Prime SP provided by Canon Medical Systems, on the real phantoms. (See Fig.~\ref{fig:ct_refmesh}) 
We used the segmentation module \cite{pinter2019polymorph} implemented in 3D Slicer \cite{fedorov20123d} for extracting the point cloud.
We first aligned the two point clouds using the iterative closest point algorithm with optimizing scale\cite{umeyama1991least} and then calculated the CD. 

The results of the skin surface recovery using hash encoding showed that the CD increased due to the noise points inside the surface.
Regarding the surface recovery of the bone, the results of the extraction using frequency encoding tended to be thicker than the extraction using the reference point group or hash encoding, and the CD value also increased accordingly.
As the resolution of the volume data from the CT is $1$ \si{mm}, it is difficult to make a precise accuracy comparison. However, the proposed method achieves surface reconstruction with no significant errors compared to the results from the helical CT.

\section{Conclusion}
We proposed NeAS, a novel method to reconstruct 3D surfaces from X-ray images and to synthesize X-ray images in new views.
NeAS leveraged a combination of signed distance functions and neural attenuation fields.
The SDF played a role in constraining the attenuation coefficient according to the boundaries of different materials.
We introduced a surface-bound function derived from SDF to adjust the learned attenuation coefficient.
Considering that camera parameters in real scenes may not always be precise, we implemented a coarse-to-fine strategy for intrinsic and extrinsic parameter refinement.
We also incorporated hash-encoding to accelerate the training and inference processes, revealing its potential in capturing high-frequency details on surfaces.
Furthermore, we demonstrated that our method provides an accessible alternative for bone density estimation using the predicted attenuation fields.

The major limitation of our method is the need for manual setting of $\alpha$ and $\beta$ to determine the ranges of the attenuation values.
Future studies should focus on developing an adaptive approach to determine these parameters. 
We also believe that expanding the proposed method to accommodate three or more materials would be beneficial. 

{\small
	\bibliographystyle{IEEEtran}
    \setlength{\itemsep}{-1pt}
	\bibliography{ref}
}

\end{document}